\documentclass[seceq]{ptptex}
\usepackage{wrapft}

\usepackage{graphicx}
\newcommand{\lk}{\left( }
\newcommand{\rk}{\right)}
\newcommand{\ltk}{\left\{ }
\newcommand{\rtk}{ \right\} }
\newcommand{\ldk}{\left[ }
\newcommand{\rdk}{ \right] }

\title{
Baryonic matter in holographic QCD
}

\author{%
Kanabu \textsc{Nawa}$^{1}$, 
Hideo \textsc{Suganuma}$^{2}$ 
and
Toru \textsc{Kojo}$^{3}$
}

\inst{
$^{1}$RCNP, Osaka University, Ibaraki, Osaka 567-0047, Japan\\
$^{2}$Department of Physics, Kyoto University, Kyoto 606-8502, Japan\\
$^{3}$RBRC, Brookhaven National Laboratory, Upton, NY11973, USA}

\abst{
We study baryons and baryonic matter in holographic QCD with
D4/D8/$\overline{\rm D8}$ multi-D brane system.
In large-$N_c$ holographic QCD, the baryon appears 
as a topologically non-trivial chiral soliton,
which is called ``brane-induced Skyrmion''. 
We also analyze the features of the baryonic matter in holographic QCD 
by investigating the system of single brane-induced Skyrmion
on a three-dimensional closed manifold $S^3$.
We propose a new interesting picture of ``pion dominance'' near the critical density.
}

\begin{document}

\maketitle

\section{Baryons in holographic QCD}

``Holography'' is a new concept of the superstring theory proposed by Maldacena 
in 1997 \cite{Mal} as the duality between a $(p+1)$-dimensional gauge theory and 
a $\{(p+1)+1\}$-dimensional supergravity, which are related through the ${\rm D}_p$ brane.
One of the most essential properties of the holography is 
the ``strong-weak duality'' between the gauge theory and the supergravity, 
and the holography provides a remarkable possibility that 
non-perturbative aspects of one side can be analyzed by 
the other dual side just with the tree-level calculations.
Then, if QCD is constructed on appropriate 
${\rm D}$ brane configurations, non-perturbative aspects of QCD
can be examined by the tree-level dual supergravity side.
This is the strategy of the holographic QCD.

In 2005, Sakai and Sugimoto succeeded in constructing QCD with 
massless quarks and gluons on the D4/D8/$\overline{\rm D8}$ 
multi-D brane configurations in type IIA superstring theory.\cite{SS}
With this model, many phenomenological properties of {\it mesons}
belonging to non-perturbative QCD are 
uniquely derived from the tree-level dual supergravity calculations.
However, since the actual classical supergravity is obtained 
for strong-coupling ``large-$N_c$'' QCD,
baryons do not directly appear as the dynamical degrees of freedom,
as a general property of large-$N_c$ QCD.\cite{tHooft}
In this sense, there occurs a problem of 
how to describe the baryons in the large-$N_c$ holographic model.

In 2006, we performed the first study of the baryon 
as a non-trivial topological soliton in holographic QCD.\cite{NSK}
This topological soliton is called as a ``brane-induced Skyrmion''.
We derive the four-dimensional meson effective action from holographic QCD 
with pions and $\rho$ mesons, 
considering the consistency with the ultra-violet cutoff scale of 
the Kaluza-Klein mass $M_{\rm KK}\sim 1{\rm GeV}$ in the holographic approach.
We obtain a stable topological soliton solution as a baryon 
in the holographic QCD,
and investigate the baryon properties (its mass, radius and energy density distribution)  
starting from the superstring theory.\cite{NSK}
(When infinite tower of the color-singlet modes are included even beyond 
the cutoff $M_{\rm KK}$, 
the soliton solution shrinks in the leading order\cite{HSSY07}.)
We find the $\rho$-meson field appearing in the core region of baryons.

\section{Baryonic matter in holographic QCD}

Now we consider an application of the holographic model to dense QCD.
According to the non-abelian nature of QCD, various interesting phases are 
expected to appear at finite temperature and density, 
and it should be important to make clear the structure of 
this ``QCD phase diagram'' based on QCD.

However, perturbative QCD breaks down at the low-energy scale 
where the QCD coupling becomes strong.
Even with lattice QCD numerical studies 
as the first principle calculation of the strong interaction, 
its applicability is severely restricted 
only near zero-density at finite-temperature in the wide QCD phase diagram, 
because of the ``sign problem''.
With these theoretical backgrounds, the holographic QCD would be 
a new important analytical tool starting from QCD to analyze the finite density regime of QCD.
This is the aim of our study. 

\begin{figure}
\begin{center}
   \resizebox{72mm}{!}{\includegraphics{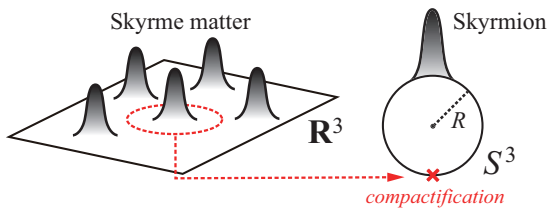}}
\caption{Schematic figure of the static Skyrme matter on a flat physical space 
${\boldmath \mbox{${\rm R}$}}^3$,
and the single Skyrmion on a closed manifold $S^3$ with a finite radius $R$.
The system of single Skyrmion on $S^3$ can be related with
the static Skyrme matter on ${\boldmath \mbox{${\rm R}$}}^3$
through the compactification of the boundary for a unit cell with single
Skyrmion on ${\boldmath \mbox{${\rm R}$}}^3$.}
\label{fig_projection1}
\end{center}
\end{figure}
 
Here, we consider the baryonic matter with large $N_c$,
because the holographic QCD is formulated as a large-$N_c$ effective theory.
In the large-$N_c$ scheme of the baryonic matter,
the kinetic energy, the N-${\rm \Delta}$ mass splitting 
and quantum fluctuations 
(apart from $O(1)$ zero-point quantum fluctuations) 
are $O(1/N_c)$, 
so that they are suppressed relative to the static baryon mass of $O(N_c)$, 
and the baryonic matter reduces to the static Skyrme matter.\cite{NSKS3}
To analyze such static Skyrme matter, we take a mathematical trick 
proposed by Manton and Ruback.\cite{MR}
To represent the high density state of the many Skyrmion system,
one unit cell shared by a Skyrmion in physical coordinate space 
${\boldmath \mbox{${\rm R}$}}^3$ 
is compactified into a three-dimensional closed manifold $S^3$
with finite radius $R$, as shown in Fig.~\ref{fig_projection1}.
The single Skyrmion placed on the surface volume $2\pi^2 R^3$ 
of the manifold $S^3$ corresponds to the finite-density baryonic matter 
with $\rho_B=1/(2\pi^2 R^3)$, and 
smaller radius $R$ of $S^3$ represents 
larger total baryon-number density of the matter.
Then, we study the baryonic matter in holographic QCD,
through the single brane-induced Skyrmion on the closed manifold $S^3$.

For the topological description of the baryon, 
we take the hedgehog configuration for the chiral (pion) field 
$U(x)=e^{i\pi(x)} \in {\rm SU(2)}_{\rm A}$ 
and the $\rho$-meson field $\rho_\mu (x)$ 
in the meson effective action derived from holographic QCD as 
\begin{eqnarray}
U^{\star}({\bf x})=e^{i\tau_a \hat{x}_a F(r)}, 
\hspace{5mm}
\rho^{\star}_{0}({\bf x})=0, \hspace{5mm} \rho^{\star}_{i}({\bf x})=\rho^{\star}_{i a}({\bf x})\frac{\tau_a}{2}
                                               =\ltk \varepsilon_{i a
                                               b}\hat{x}_b\tilde{G}(r)\rtk
                                               \tau_a,
\label{WYTP}
\end{eqnarray}
with $r \equiv |{\bf x}|$.
The pion profile $F(r)$  is a dimensionless function 
with the boundary condition of $F(0)=\pi$ and $F(\pi R)=0$ on $S^3$, 
which gives the topological charge equal to unity as a unit baryon number. 
The $\rho$-meson profile $\tilde G(r)$ has no such boundary conditions.
By the geometrical projection from the flat three-dimensional space 
${\boldmath \mbox{${\rm R}$}}^3$
onto the surface of the closed manifold $S^3$,
we obtain the energy functional of the brane-induced Skyrmion on $S^3$
for the hedgehog configuration as follows:
\begin{eqnarray}
&&\hspace{17.5mm}E[F(r), \tilde G(r)]= 
                     \int_0^{\pi R}4\pi dr R^2\sin^2\frac{r}{R}\cdot\varepsilon [F(r), \tilde G(r)], \label{BIS_energy_S3_again}\\
&&R^2\sin^2\frac{r}{R}\cdot\varepsilon [F(r), \tilde G(r)]%
          =%
             \lk R^2\sin^2\frac{r}{R}\cdot {F'}^{2}+2\sin^2F \rk
%
          +%
               \sin^2F \lk 2 {F'}^{2}+\frac{\sin^2F}{R^2\sin^2\frac{r}{R}}\rk 
               \nonumber \\ 
          &&+ 2 \lk\frac{m_{\rho}}{f_{\pi}}\rk^2%
             \ldk 4 R^2\sin^2\frac{r}{R}\cdot \tilde{G}^2\rdk
%
           - \lk 2e^2\rk g_{3\rho}  %
             \ldk 16 R\sin\frac{r}{R}\cdot \tilde{G}^3 \rdk
%
           + \lk 2e^2\rk \frac{1}{2}g_{4\rho}%
             \ldk 16 R^2\sin^2\frac{r}{R}\cdot  \tilde{G}^4 \rdk
               \nonumber \\
          &&+\lk 2e^2\rk \frac{1}{2}%
             \ldk 8\ltk
             \lk 2+\cos^2\frac{r}{R}\rk\tilde{G}^2+
             2R\sin\frac{r}{R}\cos\frac{r}{R}\cdot\tilde{G}(\tilde{G}')
             +R^2\sin^2\frac{r}{R}\cdot\tilde{G}^{'2}\rtk\rdk
               \nonumber \\
          &&+\lk 2e^2\rk  g_1 %
             \ldk 16 \ltk F'\sin F \cdot\lk
             \cos\frac{r}{R}\cdot\tilde{G}+ R\sin\frac{r}{R}\cdot \tilde{G}'\rk%
             +\sin^2F\cdot\tilde{G}/\lk R\sin\frac{r}{R}\rk\rtk\rdk
               \nonumber \\
          &&-\lk 2e^2\rk g_2%
             \ldk 16 \sin^2F\cdot\tilde{G}^2 \rdk
%
           - \lk 2e^2\rk g_3 %
             \ldk 16 \sin^2F\cdot\lk 1-\cos F\rk \tilde{G}/\lk R\sin\frac{r}{R}\rk \rdk 
               \nonumber \\
          &&-\lk 2e^2\rk g_4%
             \ldk 16 \lk 1-\cos F\rk\tilde{G}^2\rdk   
%
           + \lk 2e^2\rk g_5 %
             \ldk 16 R\sin\frac{r}{R}\cdot \lk 1-\cos F \rk \tilde{G}^3\rdk
               \nonumber  \\
          &&+\lk 2e^2\rk g_6%
             \ldk 16 R^2\sin^2\frac{r}{R}\cdot {F'}^{2}\tilde{G}^2\rdk
%
           + \lk 2e^2\rk g_7%
             \ldk 8 \lk 1-\cos F\rk^2\tilde{G}^2\rdk,\label{energy_dense_Re}
\end{eqnarray}
where $F'\equiv\frac{dF(r)}{dr}$ 
and      
$\tilde{G}'\equiv\frac{d\tilde{G}(r)}{dr}$,
and we use Adkins-Nappi-Witten (ANW) unit\cite{ANW}
for the unit of length and energy. 
All the coupling constants 
($e$, $g_{3\rho}$, $g_{4\rho}$, $g_{1\sim 7}$) in Eq.(\ref{energy_dense_Re}) are uniquely
determined by just two experimental inputs,
$f_\pi=92.4\mbox{MeV}$ and $m_\rho=776\mbox{MeV}$.
Such uniqueness is one of the remarkable
consequences of the holographic framework.

\begin{figure}[t]
  \begin{center}
       \begin{tabular}{cc}
\resizebox{68.5mm}{!}{\includegraphics{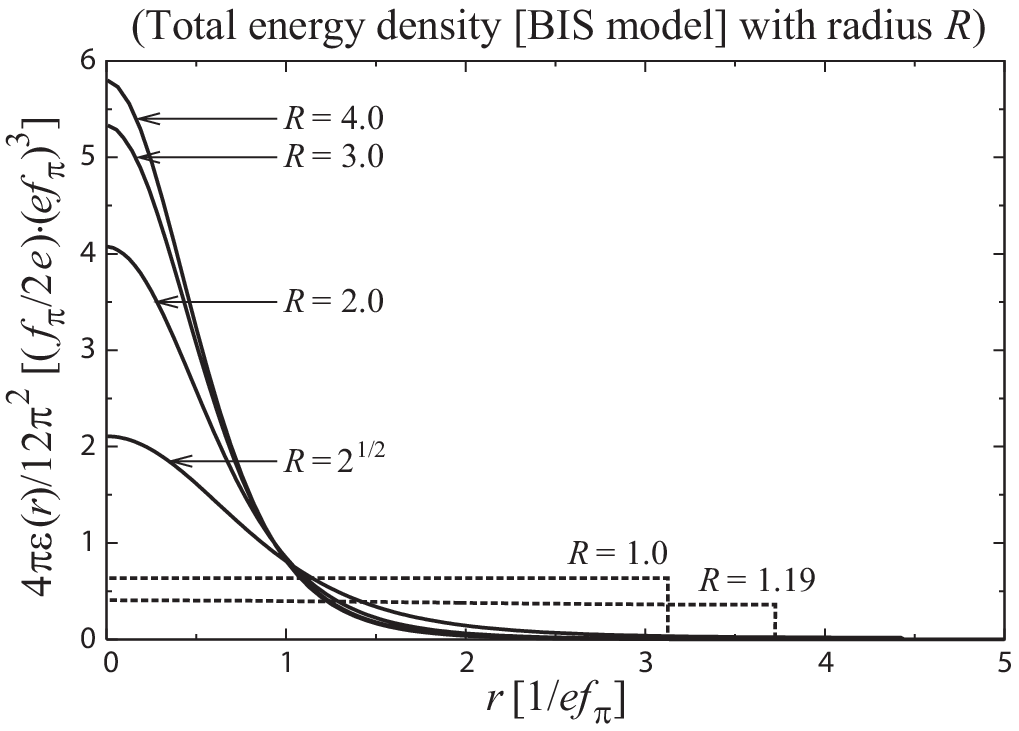}}
&
       \hspace{-4mm}
%
\resizebox{68.5mm}{!}{\includegraphics{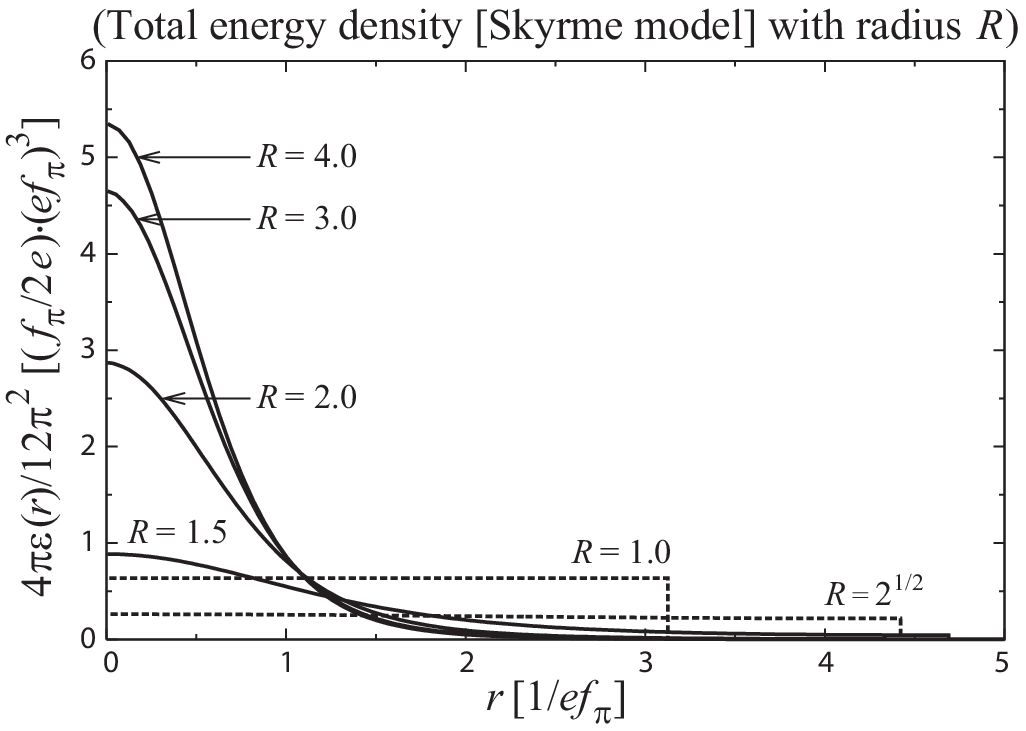}}
       \end{tabular}
  \end{center}
\caption{The total energy density of single baryon 
on $S^3$ with various values of radius $R$ in ANW unit: (left) 
the brane-induced Skyrme (BIS) model;
(right) the Skyrme model.
}
  \label{fig_conf1}
\end{figure}

Solving the field equations with the topological boundary condition 
of $F(0)=\pi$ and $F(\pi R)=0$,
we obtain the pion profile $F(r)$ and the $\rho$-meson profile $\tilde{G}(r)$
as the hedgehog soliton solution for the energy functional (\ref{BIS_energy_S3_again}) at each $R$.
In Fig.~\ref{fig_conf1}, we show the $R$-dependence of 
the total energy density of single Skyrmion on the manifold $S^3$ for 
the brane-induced Skyrme (BIS) model, together with the result in 
the Skyrme model (without $\rho$ mesons).
As the radius $R$ of $S^3$ decreases, i.e.,
as the baryon number density $\rho_B=1/(2\pi^2 R^3)$ increases, there 
appears delocalization of the energy density of the baryon in both models.
We find the ``delocalization phase transition''
into the uniform phase 
at $R=R_{\rm crit}^{\rm BIS}=1.19$
for the BIS model and 
at $R=R_{\rm crit}^{\rm Skyrme}=\sqrt{2}$
for the Skyrme model.
Such delocalization phase transition is expected to relate to 
the deconfinement in the presence of baryons. 
It is also related to the chiral restoration in the bulk hadronic matter
in terms of the spatially-averaged chiral condensate
as the global order parameter of the chiral symmetry~\cite{NSKS3,Fo}.

Taking the two experimental inputs,
$f_\pi=92.4\mbox{MeV}$ and $m_\rho=776\mbox{MeV}$,
we get the critical density of the delocalization phase transition
as $7.12 \rho_0$ for the BIS model and 
$4.26 \rho_0$ for the Skyrme model.\cite{NSKS3}
For the BIS model, the heavy $\rho$ meson appearing in the core region of 
the baryon is to provide the attraction with the pion field, 
which leads to the shrinkage of 
the total size of the baryon~\cite{NSK}.
Owing to the shrinkage of the baryon due to $\rho$ mesons, larger baryon number density 
is needed for the BIS model to give the delocalization phase transition.

\begin{figure}
   \resizebox{140mm}{!}{\includegraphics{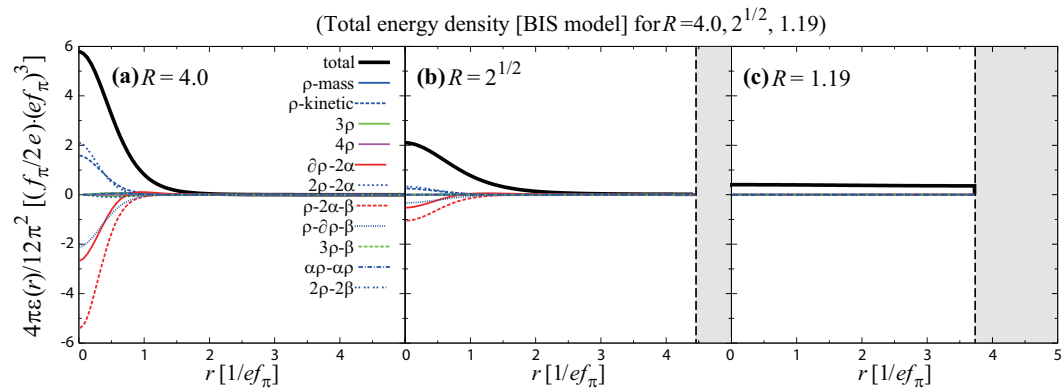}}
\caption{The total energy density and each contribution from 
the $\rho$-meson interaction terms in Eq.(\ref{energy_dense_Re}) 
for the single brane-induced Skyrmion on $S^3$ with the radius 
(a) $R=4.0$, (b) $R=\sqrt{2}$, and (c) $R=1.19$ in ANW unit. 
The vertical dashed line in (b) and (c) 
denotes $r=\pi R$, the maximal value on $S^3$.
For $R=R_{\rm crit}^{\rm BIS}=1.19$, the total energy density becomes 
a uniform distribution as the identity map,
where the $\rho$-meson field disappears and only the pion field survives, 
indicating ``pion dominance''.}
\label{fig_rhocont}
\end{figure}

In Fig.\ref{fig_rhocont}, we compare 
the total energy density of the brane-induced Skyrmion and 
each contribution from the $\rho$-meson interaction terms 
in Eq.(\ref{energy_dense_Re}) for $R=4.0$, $\sqrt{2}$, and 1.19.
For $R=4.0$ and $\sqrt{2}$, the $\rho$-meson field 
appears in the core region of the baryon.
On the other hand, at $R=R_{\rm crit}^{\rm BIS}=1.19$,
the $\rho$-meson field and its contributions disappear in the
uniform phase.

We conjecture that such disappearance of the $\rho$-meson field 
in high density phase
around the critical density can be generalized to all the other (axial) vector mesons 
even including the heavier mesons, $a_1, \rho', a'_{1}, \rho''\cdots$,
denoted by the field $B_\mu^{(n)}(x_\nu)$ with the mass $m_n$ 
($m_1 < m_2 < ...$), 
by the following reasons:
\begin{enumerate}
\item[1)]
The kinetic term of the (axial) vector meson field $B_\mu^{(n)}$
appears on $S^3$ as 
${\rm tr}\{\partial_\mu B_\nu^{(n)}-\partial_\nu
B_\mu^{(n)}\}^2\propto R^{-2}$.
Therefore, the spatial variation of the field $B_\mu^{(n)}(x)$
is suppressed for small $R$, i.e., at high density.
\item[2)]
The mass term $m_n^2 {\rm tr}\{ {B_\mu^{(n)}B_\mu^{(n)}}\}$ suppresses 
the absolute value of the $B_\mu^{(n)}$ field because of its large mass $m_n^2$.
\item[3)]
The coupling between pions and heavier (axial) vector mesons
$B_\mu^{(n)}$ with larger index $n$ 
is found to become smaller both in the phenomenological 
dimensional deconstruction model\cite{DTSon} and in 
the holographic QCD.\cite{NSK}
Hence, the effect of heavier (axial) vector mesons
becomes smaller for the baryon as the chiral soliton, which basically 
consists of a large-amplitude pion field.
\end{enumerate}
These considerations 1), 2), 3) would support our conjecture.
In other words, only the pion field survives in the baryonic matter 
near the critical density, which we call ``pion dominance''.
The effect of (axial) vector mesons only appears through their interaction 
with the pion field, which affects the actual value of the critical density.

Note that the pion field 
cannot disappear because of the boundary condition, 
$F(0)=\pi$ and $F(\pi R)=0$ to maintain the unit baryon number 
on each closed manifold $S^3$.
On the other hand, there is no constraint for the (axial)
vector meson fields, and they can disappear in high density phase.
Such pion dominance would be somehow related with the importance of chiral soft modes 
in the strong-coupling QGP (sQGP) at finite temperature in Refs. 11) and 12).

In summary, we have studied baryons and baryonic matter in holographic QCD.
We have investigated the baryonic matter in terms of 
single brane-induced Skyrmion on a three-dimensional closed manifold $S^3$, 
and have found a new interesting picture of ``pion dominance'' near the critical density. 

\vspace{0.5cm}
\noindent
{\bf Note Added:} After completing this study, we noticed a related paper,\cite{KSZ} 
where the authors used instantons on the D8 brane
in the Wigner-Seitz approximation.

\vspace{0.5cm}
\noindent
{\bf Acknowledgement:} We thank YITP at Kyoto University 
for useful discussions 
during the Int. Symp. on ``Fundamental Problems in Hot and/or Dense QCD''.

\end{document}